\documentstyle[ws-procs975x65]{article}

\input psfig.sty

\def\be{\begin{equation}}
\def\ee{\end{equation}}
\def\bea{\begin{eqnarray}}
\def\eea{\end{eqnarray}}

\def\li11{$^{11}Li$}
\def\o18{$^{18}O$}

\begin{document}


\title{{\Large {\bf Dipole Symmetry Near Threshold}} \footnote{Work Supported by USDOE Grant 
   No. DE-FG02-94ER40870.}}

\author{\Large Moshe Gai}

\address{Laboratory for Nuclear Science, Dept. of Physics, University of Connecticut, \\
2152 Hillside Rd, U3046, Storrs, CT 06269-3046, USA \\ 
   gai@uconnvm.uconn.edu \ \ - \ \ http://www.phys.uconn.edu}


\maketitle

\abstracts{In celebrating Iachello's 60th birthday we underline  
many seminal contributions for the study of the degrees of freddom relevant 
for the structure of nuclei and other hadrons. A dipole degree of freedom, 
well described by the spectrum generating algebra U(4) and the Vibron Model, 
is a most natural concept in molecular physics. It has been suggested by 
Iachello with much debate, to be most important for understanding the low 
lying structure of nuclei and other hadrons. After its first observation in
$^{18}O$ it was also shown to be relevant for the structure of heavy nuclei (e.g. 
$^{218}Ra$). Much like the Ar-benzene molecule, it is shown that molecular configurations 
are important near threshold as exhibited by states with a large halo and strong 
electric dipole transitions. The cluster-molecular Sum Rule derived by Alhassid, 
Gai and Bertsch (AGB) is shown to be a very useful model independent tool for 
examining such dipole molecular structure near thereshold. Accordingly, the 
dipole strength observed in the halo nuclei such as $^6He, \ ^{11}Li, \ ^{11}Be, 
^{17}O$, as well as the N=82 isotones is concentrated around threshold 
and it exhausts a large fraction (close to 100\%) of the AGB sum rule, but a 
small fraction (a few percent) of the TRK sum rule. This is suggested as 
an evidence for a new soft dipole Vibron like oscillations in nuclei.}

\section{Molecular Dipole Symmetry}

A molecular degree of freedom is characterized by excitations that 
involves the relative motion of two tightly bound constituents and not 
the excitation of the objects themselve. Hence it is associated 
with a polarization vector known as the separation vector. Such a 
vector can be classicaly described in a geometrical model in three 
dimensions or by using the corresponding group U(4) \cite{Ia82} and 
the very succesful Vibron model of molecular Physics \cite{Levine}. This 
model has two symmetry limits that correspond to the geometrical description 
of Rigid Molecules, the O(4) limit, or Soft Molecules, the U(3) limit.

A most comprehensive discussion of such molecular structure and the 
Vibron model can be found in Iachello-Levine book \cite{Levine} on 
"Algebraic Thoery of Molecules".  In Fig. 1 taken from that book we show 
the characteristic dimensions of the Ar-benzen molecule. The argon atom 
is losely bound to the (tightly bound) benzen molecule by a van der Waalls 
polarization and thus this molecular state lies  close to the dissociation 
limit. We note that the relative dimension and indeed the very polarization 
phenomena are reminscent of a halo structure where the argon atom creates 
a "halo" around the benzen molecule.

\centerline{\psfig{figure=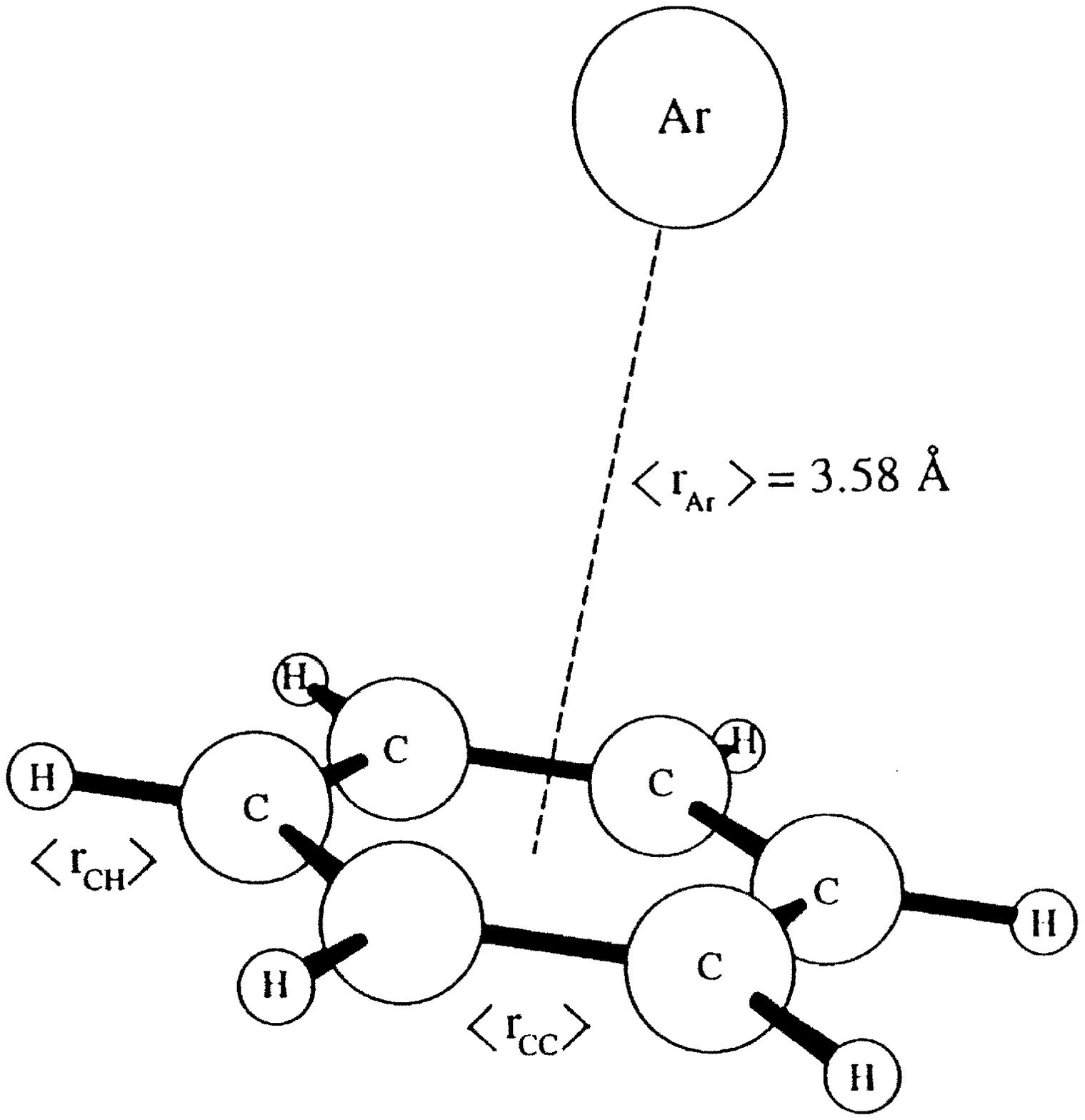,height=2.5in}}

\underline{Fig. 1:}  Characterstics dimensions of the Ar-benzen 
  molecule, adopted from Iachello and Levine \cite{Levine}.

\section{The AGB Cluster Sum Rule}

The polarization phenomena associated with a molecular state 
implies that it should be associated with dipole excitations of the 
separation vector. In this case expectation values of the dipole 
operator do not vanish as the center of mass and center of charge 
of the polarized molecular state do not coincide \cite{Rad,Gell}.
Hence  molecular states give rise to low lying dipole excitations. While 
the high lying Giant Dipole Resonace (GDR) is associated with a 
Goldhaber-Teller \cite{Gold} excitation of the entire neutron distribution 
against the proton distribution, a molecular excitation involves a smaller fraction 
of the nucleus at the surface and is expected to occur at lower 
excitation than the GDR; i.e. a soft dipole mode \cite{Ik,Hans}.

The GDR exhausts the Thomas-Reiche-Kuhn (TRK) \cite{TRK}
Energy Weighted Dipole Sum Rule as applied to nuclei:
\   \\

\begin{tabbing}
 \hspace{0.5in} $S_1(E1;A)$ \= 
    = $\Sigma_i \ B(E1:0^+ \ \rightarrow \ 1^-_i) \times E^*(1^-_i)$ \\
        \> = ${9 \over 4 \pi} \ {NZ \over A} \times {e^2 \hbar ^2 \over 2m}$
       \hspace{1.9in} \= (equ. 1) \\
\end{tabbing}
And for a molecular state Alhassid, Gai and Bertsch \cite{AGB} 
derived sum rules by subtracting the individual 
sum rules of the contituents from the total sum rule:
\   \\

\begin{tabbing}
\hspace{0.1in} $S_1(E1;A_1 \ + \ A_2)$ \= 
   = $S_1(A) \ - \ S_1(A_1) \ - \ S_1(A_2)$ \\
\> = ${9 \over 4 \pi} \ {{(Z_1A_2 \ - Z_2A_1)^2} \over AA_1A_2} \times {e^2 \hbar ^2 \over 2m}$
    \hspace{1.2in} \= (equ. 2) \\
\   \\
\hspace{0.1in} $S_1(E1;\alpha+A_2)$ \> = ${9 \over 4 \pi} \ {(N-Z)^2 \over A(A-4)} 
 \times {e^2 \hbar ^2 \over 2m} $ \> (equ. 3) \\
\hspace{0.1in} $S_1(E1;n+A_2)$ \> = ${9 \over 4 \pi} \ {Z^2 \over A(A-1)} 
 \times {e^2 \hbar ^2 \over 2m} $ \> (equ. 4) \\
\hspace{0.1in} $S_1(E1;2n+A_2)$ \> = ${9 \over 4 \pi} \ {2Z^2 \over A(A-2)} 
 \times {e^2 \hbar ^2 \over 2m}$ 
\> (equ. 5) 
\end{tabbing}

The molecular sum rule, equ (2), was shown to be useful in elucidating 
molecular (cluster) states in $^{18}O$ where the measured B(E1)'s and 
B(E2)'s exhaust 13\% and 23\%, respectively, of the molecular sum rule
\cite{Ga83}. Similarily, these molecular states in $^{18}O$ 
have alpha widths that exhaust 20\% of the Wigner sum rule. The 
branching ratios for electromagnetic decays in $^{18}O$ were also 
shown to be consistent with predictions of the 
Vibron model in the U(3) limit \cite{Ga91}. Indeed the manifestation 
of a molecular structure in $^{18}O$ has altered our 
undertsanding of the coexistence of degrees of freedoms 
in $^{18}O$ \cite{Ga89}. Similar observations were also 
made in the heavy nucleus $^{218}Ra$ \cite{Ra}.

\centerline{\psfig{figure=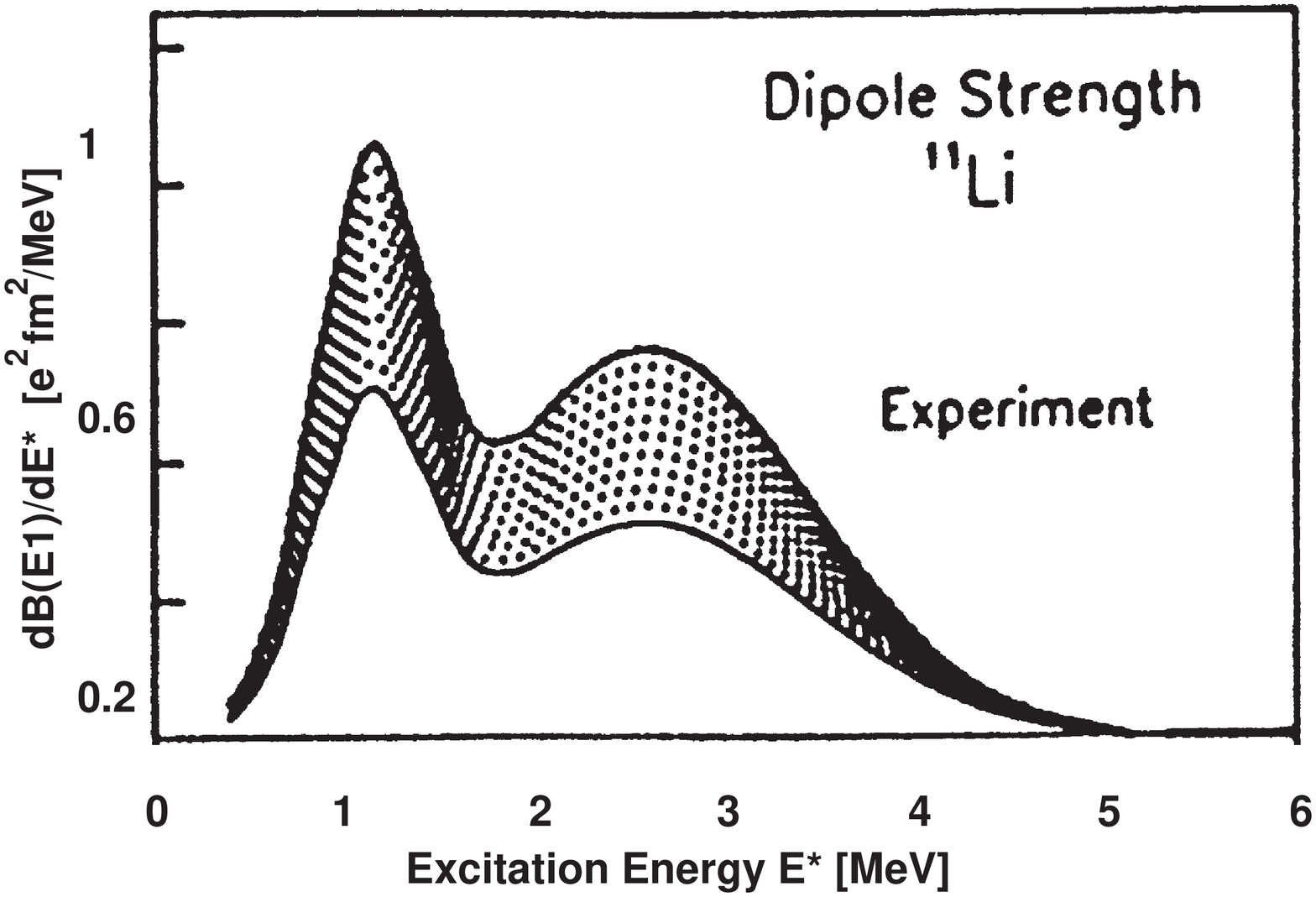,height=2.5in}}

\underline{Fig. 2:}  Dipole strength measured in \li11 \cite{Zins}.

\centerline{\psfig{figure=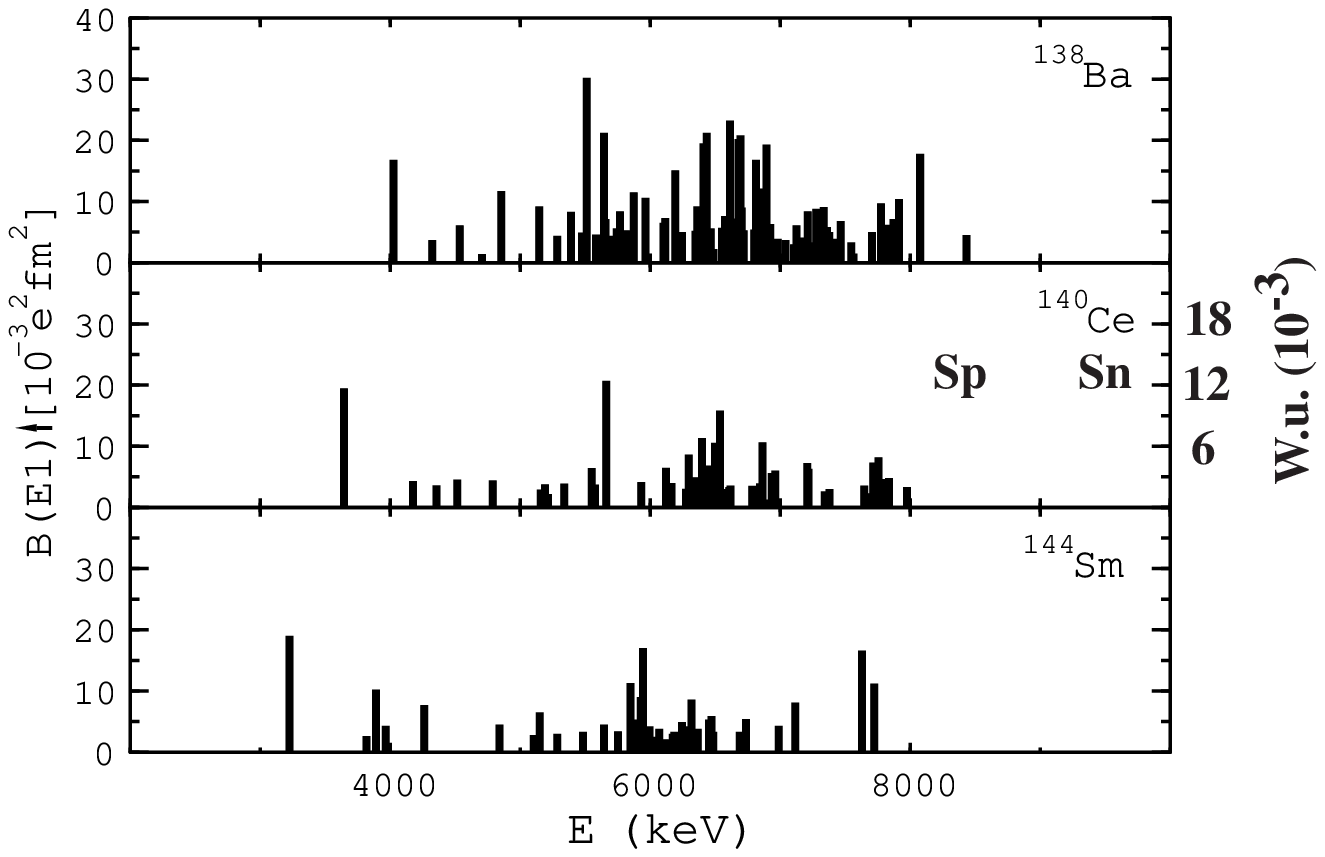,height=2.5in}}

\underline{Fig. 3:}  Dipole strength measured in N=82 isotones \cite{Zilg}.

The dipole strength at approximately 1.2 MeV in \li11 \cite{Zins}, 
shown in Fig. 2, exhausts approximately 20\% of the molecular sum rule, and 
the total strength integrated up to 5 MeV exhausts approximately 100\% of the 
cluster sum rule \cite{Fox,Gai99}, but it only exhausts approximately 8\% of 
the TRK sum rule, see table 1. We emphasize that the experimental efficiency at 
for example 6 MeV is very large (30\%), but no strength is found at higher 
energies beyond 100\% of the molecular sum rule. These two facts strongly 
suggest the existence of a low lying soft dipole mode in \li11. Similar 
observation are reported in $^{11}Be$ \cite{Be11}, oxygen isotopes
\cite{Oxygen} and $^6He$ \cite{He6}, believed to exhibit a halo structure. 
The N=82 isotones also show a diople strength near threshol as shown 
in Fig 3 \cite{Zilg}. These results are summarized in Table 1.

The ratio of the TRK/AGB sum rules is given by:
\   \\

 \begin{tabbing}

\hspace{1in} TRK/AGB  \= 
   = $NZA_1A_2/(Z_1A_2 \ - \ Z_2A_1)^2$ \ \ \ \ \ \ \ \ (equ. 6) \\
\hspace{1.2in} ($\alpha)$ \> = $(N-Z)^2/NZ(A-4)$ \\
\hspace{1.2in} (1n) \> = N(A-1)/Z \\ 
\hspace{1.2in} (2n) \> = N(A-2)/2Z \\

\end{tabbing}

\begin{table}[hb] 


\caption[]{Measured E1 strength in nuclei.}\label{tab1}


\begin{center}

\begin{tabular}{lllll}

\hline\\[-10pt]
Nucleus & $<E^*>$ & TRK & TRK/AGB & AGB \\
\hline\\[-10pt]
$^{11}Li$ \cite{Zins,Fox} & 1.2 MeV & $8.0 \pm 2.0$\% & (2n) 12 & $96 \pm 24$\%  \\
$^{11}Be$ \cite{Be11} & 1.0 MeV & 5.0\% & (1n) 18 & 90\%  \\
$^{17}O$ \cite{Oxygen} & $<$ 15 MeV & 4\% & (1n) 18 & 72\%  \\
$^{138}Ba$ \cite{Zilg} & 6.5 MeV & $0.78 \pm 0.15$\% & (1n) 200 & $156 \pm 30$\%  \\
{\phantom{$00$}}\\
\hline 
\end{tabular}
\end{center}
\end{table}

\section{Conclusions}

In conclusions we demonstrate that molecular configurations play a major 
role in the structure of light and heavy nuclei.  Unlike the Giant Dipole 
Resonance that involves oscillation of the entire neutron-proton distributions, 
these Vibron states involve only oscillations of the surface of the nucleus, and 
hence they lie at lower energies than the GDR. Similarly, while the GDR 
exhausts the TRK sum rule, the Vibron states exhausts the ABG cluster 
sum rule.


\begin{thebibliography}{99}

\bibitem{Ia82} F. Iachello, and A.D. Jackson; 
     Phys. Lett. {\bf 108B}(1982)151.

\bibitem{Levine} F. Iachello and R.D. Levine, Algebraic Theory 
   of Molecules; Oxford University Press, 1995.

\bibitem{Rad} L.A. Radicati; Phys. Rev. {\bf 87}(1952)521.

\bibitem{Gell} M. Gell-Mann and V.L. Telegdi; Phys. Rev. {\bf 91}(1953)169.

\bibitem{Gold} M. Goldhaber and E. Teller; Phys. Rev. {\bf 74}(1948)1046.

\bibitem{Ik} K. Ikeda Nucl. Phys. {\bf A538}(1992)355c.

\bibitem{Hans} P.G. Hansen; Nucl. Phys. {\bf A588}(1995)1c. \\ P.G. Hansen 
    and A.S. Jensen; Annu. Rev. Nucl. Part. Sci. {\bf 45}(1995)591.

\bibitem{TRK} W. Kuhn; Zeit. f. Phys. {\bf 33}(1925)408. F. Reiche, W. 
  Thomas; Zeit. f. Phys. {\bf 34}(1925)510.

\bibitem{AGB} Y. Alhassid, M. Gai, and G.F. Bertsch ; 
     Phys. Rev. Lett. {\bf 49}(1982)1482.

\bibitem{Ga83} M. Gai, M. Ruscev, A.C. Hayes, J.F. Ennis, R. Keddy, 
   E.C. Schloemer, S.M. Sterbenz and D.A. Bromley; 
   Phys. Rev. Lett. {\bf 50}(1983)239.

\bibitem{Ga91} M. Gai {\em et al.}; Phys. Rev. {\bf C43}(1991)2127.

\bibitem{Ga89} M. Gai {\em et al.}; Phys. Rev. Lett. {\bf 62}(1989)874.

\bibitem{Ra} M. Gai {\em et al.}; Phys. Rev. Lett. {\bf 51}(1983)646.

\bibitem{Zins} M. Zinser {\em et al.}; Nucl. Phys. {\bf A619}(1997)151.

\bibitem{Fox} G.F. Bertsch and J. Foxwell; Phys. Rev. {\bf C41}(1990)1300.

\bibitem{Gai99} M. Gai; Rev. Mex. Fis. Supp. {\bf 45}(1999)106.

\bibitem{Be11} T. Nakamura {\em et al.}; Phys. Lett. {\bf B331}(1994)296.
 N. Gan {\em et al.} http://www.phy.ornl.gov/progress/ribphys/reaction/rib023.pdf.

\bibitem{Oxygen} T. Aumann {\em et al.}; Nucl. Phys. {\bf A649}(1999)297c.
  A. Leistenscheneider {\em et al.}; Acta. Phys. Pol. {\bf B32}(2001)1095.

\bibitem{He6} S. Nakayama {\em et al.}; Phys. Rev. Lett. {\bf 85}(2000)262.

\bibitem{Zilg} A. Zilges {\em et al.} Phys. Lett. {\bf B542}(2002)43. 

\end{thebibliography}
\end{document}